# Electronic Spectroscopy of Atomic Defects in Molybdenum Disulfide under Ambient Conditions

Joshua R. Evans[1], Diego A. Garibay[2], Aiden N. Kuhls[2], and Mehmet Z. Baykara[2,a)]

[1] Department of Materials Science and Engineering, University of California Merced, Merced, CA 95343, USA

[2] Department of Mechanical and Aerospace Engineering, University of California Merced, Merced, CA 95343, USA

Transition metal dichalcogenides (TMDs) attract significant attention as potential building blocks in next-generation electronic devices. On the other hand, a comprehensive understanding of how various defects affect local electronic properties under realistic operational conditions is yet to be formed. Here, we present results of electronic spectroscopy experiments performed on individual defects in the prototypical TMD molybdenum disulfide ($MoS_2$) under ambient conditions, by way of conductive atomic force microscopy (C-AFM). Data acquired in the form of consecutive, high-resolution current maps at various bias voltages allow the assessment of local conductivity and differential conductance as a function of bias voltage for individual defects, the effects of which range from single atomic sites to several nanometers in lateral size. Characteristic behavior in spectroscopy data allows the categorization of observed defects into distinct groups, and their chemical identification as n-type and p-type transition metal substitutions of molybdenum atoms, as well as oxygen substitutions of sulfur atoms.

[a)] Author to whom correspondence should be addressed: mehmet.baykara@ucmerced.edu

Two-dimensional (2D) materials are being intensely studied since the discovery of the outstanding electrical properties exhibited by graphene in 2004.[1] 2D materials are attractive for a wide range of applications due to geometric reasons; their extremely small thickness facilitates device designs (such as those associated with transistors), where the ability to fabricate multiple components in small spaces becomes critically important. In addition to the reduction of size that they enable, 2D materials exhibit attractive physical properties in terms of mechanics, electronics, and chemistry. One important class of 2D materials is transition metal dichalcogenides (TMDs), which possess intriguing optoelectronic characteristics (e.g., direct and tunable bandgaps) which make them promising candidates for key components in next-generation electronic devices.[2]

An important factor (and a potential obstacle) to the successful implementation of 2D TMDs in next-generation electronic devices is the unavoidable presence of atomic-scale defects in these materials, which may include atomic vacancies, adatoms, atom substitutions, and clusters thereof.[3] While certain atomic defects exist in TMDs naturally,[4] introduction of atomic defects may also occur during laboratory synthesis of the materials.[5] On the one hand, defects can significantly reduce carrier mobility by acting as local scattering sources.[6] Defects were also shown to modify the bandgap in 2D TMDs.[7] In particular, theory and experiments have shown that certain atomic defects can disturb local electronics by introducing in-gap states (electronic states that arise in the otherwise unpopulated bandgap of the 2D TMD), leading to an overall increase in electronic heterogeneity with potentially unfavorable effects on device performance.[8] Lastly, the uneven (i.e., heterogenous) spatial distribution of atomic defects in 2D TMDs has the potential to lead to inconsistencies and reliability concerns in nanoscale device batches based on these materials.[9]

The effects outlined above highlight the need for precise defect characterization (both in terms of atomic structure and electronic properties) in 2D TMDs as a crucial step facilitating their implementation in device applications. Moreover, such information can help refine synthesis processes to control the types and densities of atomic defects in lab-grown 2D TMDs.[10] Multiple characterization techniques exist with which atomic defects can be probed in 2D TMDs. Specifically, scanning probe microscopy (SPM) methods have shown promising results with their ability to scan local regions of materials with high spatial resolution. Among SPM methods used towards characterizing atomic defects, scanning tunneling microscopy (STM) is the most conventional one, together with scanning tunneling spectroscopy (STS) that reveals the electronic density of states (DOS) associated with defective sites, facilitating defect identification based on characteristic changes in DOS.[5, 11]

A critical limitation associated with STM is the requirement of performing experiments under ultrahigh vacuum (UHV) conditions, due to the necessity of molecularly clean surfaces. This leads to various complications, including low throughput due to extended sample preparation and measurement procedures. Perhaps equally importantly, the UHV environment is not representative of the typical conditions under which devices based on 2D TMDs operate, potentially limiting the relevance of results attained under UHV conditions for practical applications.

In recent years, atomic-resolution imaging was shown to be achievable under ambient conditions via conductive atomic force microscopy (C-AFM), on various 2D and bulk materials.[12, 13] Although the underlying physical mechanisms are still under debate, it was demonstrated that C-AFM can image, on 2D TMDs, the same atomic defects as STM.[14] The use of C-AFM for characterizing defect densities in 2D TMDs has also been discussed.[9] On the other hand, C-AFM images (much like other SPM images) alone do not provide direct information on the chemical identity of the defects, complicating data interpretation and preventing the formation of direct connections between fundamental results and implications on the device scale. This inability to identify defects limits the full potential of C-AFM for atomic defect characterization on 2D TMDs under ambient conditions. It should be noted that while measurements of current ($I$) as a function of bias voltage ($V$), reminiscent of STS, may indeed be performed by C-AFM,[15, 16] this approach is not suitable for studying the electrical characteristics of atomic defects due to problems associated with thermal drift under ambient conditions. In particular, lateral misalignments between the probe and the tip during data collection result in significantly reduced levels of confidence in assigning recorded DOS features to specific, atomic-scale defects.

Motivated in this fashion, we present here a C-AFM-based approach termed "discrete $I$-$V$ spectroscopy" which enables the unambiguous electronic spectroscopy of individual atomic defects under ambient conditions. Our approach, applied on the prototypical TMD molybdenum disulfide ($MoS_2$), allows us to categorize observed defects into various groups based on their bias-voltage-dependent electronic behavior, and provides important information with regards to their identities by comparison with STS and density functional theory (DFT) data in the literature.

For the experiments presented here, bulk, flux-grown $MoS_2$ crystals were obtained commercially (*2D Semiconductors*) and mechanically exfoliated via the adhesive tape method[1] onto conductive substrates in the form of gold-coated silicon wafers (*Ted Pella*). Conductive substrates are necessary to achieve imaging via C-AFM, as they provide a conductive path to ground, whereas insulating substrates (such as mica) do not. The C-AFM imaging and spectroscopy experiments were performed with a commercial instrument (Cypher VRS, *Asylum Research, Oxford Instruments*). Conductive-diamond-coated tips were previously found to be the best probes for achieving atomic-resolution imaging in C-AFM, including the imaging of defects on or near the sample surface.[12] Specifically, CDT-CONTR probes (*Nanosensors*) are used in our work, along with AD-2.8-AS (*Adama*) probes. In our experiments, atomic resolution is best achieved using high-speed scanning (at a scanning frequency of 15.62 Hz, typically corresponding to scanning speeds above 100 nm/s), which counteracts the effects of thermal drift, and was additionally hypothesized to hinder enlarged electrical contact areas over the water meniscus at the tip-sample junction.[17] Experiments are performed under ambient conditions (typical temperature range: 22−23 °C; typical relative humidity range: 20−50%).

Fig. 1 demonstrates the capability of C-AFM to image defects of various size on exfoliated $MoS_2$ flakes under ambient conditions. Specifically, Fig. 1(a) comprises a $10 \times 10$ $nm^2$, atomic-resolution current image where a point (i.e. single-atomic) defect can be clearly observed, locally enhancing the conductivity. Such single-atomic, current-enhancing defects were previously observed in $MoS_2$ and attributed to sub-surface / buried substitutions of a S atom, potentially by O.[5, 18] The capability to image single-atomic defects fulfills the criterion of *true* atomic-resolution imaging, with images of similar quality to what has been presented before in the literature.[12] The $10 \times 10$ $nm^2$ current image presented in Fig. 1(b) complements the imaging of a point defect in Fig. 1(a). More specifically, a current-enhancing defect spanning multiple atomic sites over a lateral distance of ~1.5 nm is observed in Fig. 1(b), together with the underlying atomic lattice of the $MoS_2$ surface. It is important to note that the point defect in Fig. 1(a) proves useful for characterizing the lateral thermal drift between the tip and the sample during imaging experiments. By precisely following the position of the point defect across 7 scans recorded subsequently over a period of approximately 1.8 minutes, the drift rate was estimated to be ~6 Å/min laterally, which is sufficiently low for recording multiple current images focusing on the same defect (or a group of defects) over extended periods of time.

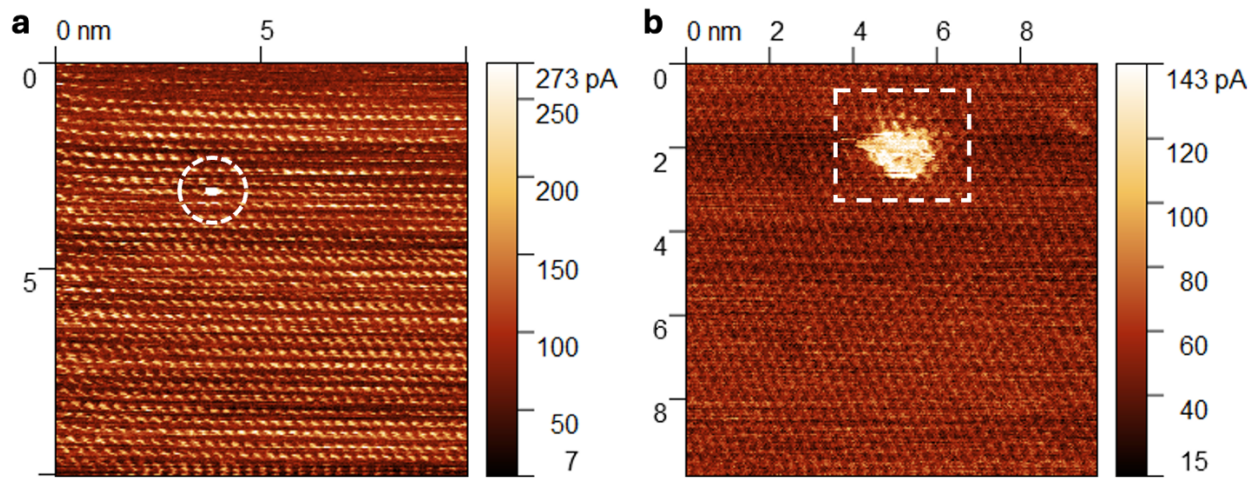

**FIG 1.** (a) Current map comprising a point defect (highlighted by the dashed white circle) encountered on the MoS$_2$ surface, recorded at a bias voltage of +2.5 V. (b) Current image showing a larger defect spanning multiple lattice sites (highlighted by the dashed white rectangle), recorded at a bias voltage of +2 V.

While relatively small current images (such as those presented in Fig. 1) are useful for characterizing the atomic structure of individual defects, larger images are required to make reasonable conclusions about the variety and spatial density of defects encountered in 2D TMDs such as MoS$_2$ — important criteria that are expected to affect device-level performance. As such, additional images were taken on the MoS$_2$ surface at larger scan sizes (from several tens of nm across, up to 150 × 150 nm$^2$), which revealed the presence of various defects that perturb local conductivity (Fig. 2). Defects were found to differ in terms of their lateral size, as well as their spatial morphology, and notably, their bias-voltage-dependent electronic behavior. The latter point is illustrated by pairs of current images recorded over the same area at two different bias voltages (Fig. 2(a, b) and Fig. 2(c, d)). While some defects appear to enhance conductivity at both positive and negative bias voltages, others only appear (i.e., are detectable in terms of their influence on the local conductivity) at one bias polarity. Moreover, the spatial morphology of the electronic perturbation induced by the defects appear to change at different bias polarities, as clearly observed in Fig. 2(c, d). The fact that different defects exhibit distinct morphological features in a single scan frame rule out artificial factors such as tip asymmetry and double tip apexes influencing the data. These observations thus open the avenue of using voltage-dependent electronic behavior as the "fingerprint" for identifying defects in 2D TMDs such as $MoS_2$ under ambient conditions.

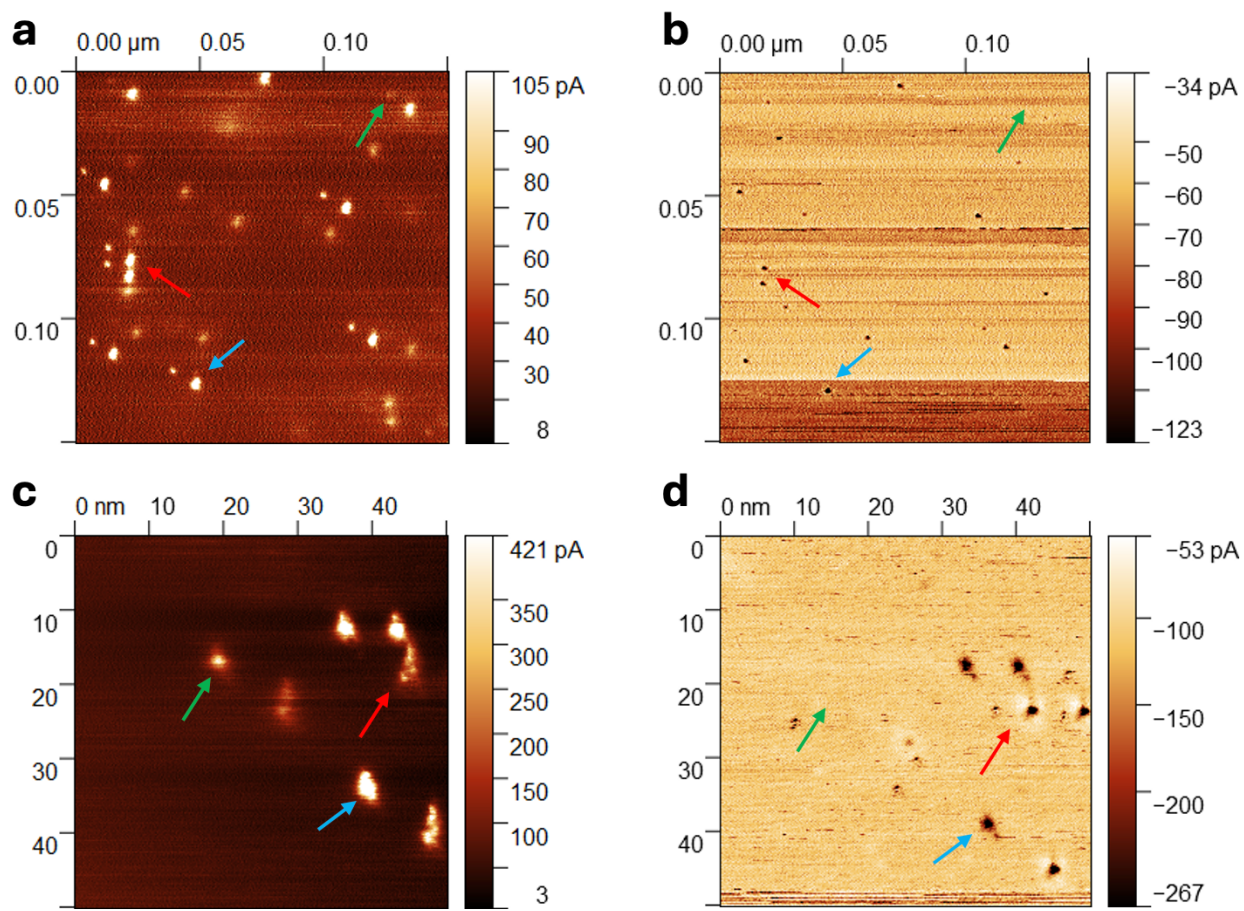

**FIG 2.** Pairs of current maps recorded at different bias voltages on $MoS_2$. The top row shows two 150 × 150 nm² current images on the same area, recorded at two different bias voltages: (a) +2 V, and (b) -2 V. Note that while some defects are enhancing current at both bias polarities (blue and red arrows), others are not detectable at negative bias voltage (green arrow). The bottom row shows two 50 × 50 nm² current images, again recorded at (c) +2 V and (d) -2 V. The different spatial

morphologies exhibited by defects at opposing polarities are clear (blue and red arrows), while a particular defect is again not detectable at negative bias voltage (green arrow). There is a slight lateral shift between the images in (c) and (d) induced by thermal drift, which results in an additional defect entering the scanning frame on the right side in (d).

In order to study the electronic behavior of individual defects, electronic spectroscopy needs to be performed in the form of *I-V* curves, from which d*I*/d*V*-*V* (differential conductance, which is proportional to electronic DOS) behavior can be extracted. To overcome the thermal drift issue with conventional *I-V* spectroscopy discussed above, we developed a new methodology for the collection of bias-voltage-dependent current. In particular, instead of placing the probe over a particular defect and applying a sweeping bias voltage, the new technique operates by recording sets of multiple, complete current images on an area with the defect(s) of interest at discrete bias voltages (e.g., 5 scans at +2 V, followed by 5 scans at +1.5 V, and so on, until the ultimate value of -2 V is reached). This process, named "discrete *I-V* spectroscopy" partially mimics the continuous sweeping of bias voltage values in conventional *I-V* spectroscopy by recording complete images at discrete voltage values. While the spread of data points in terms of bias voltage is significantly less dense compared to conventional *I-V* spectroscopy, notable benefits of the new technique are the following: (i) no uncertainty in assigning the recorded data to specific locations on the surface, (ii) the ability to characterize multiple defects simultaneously as they are imaged within the same scanning frame (as in Fig. 2) and (iii) the ability to track the morphology of electronic perturbations induced by the defects in real space.

In order to assign specific current values to individual defects in a given current map, we fit a 2D Gaussian surface to an elliptical area around the defect, whereby the amplitude of the Gaussian surface corresponds to the current value associated with that defect. This procedure is repeated for all (i.e., 5) images recorded at a given bias voltage; the average amplitude value is then assigned to that defect as its characteristic current at that voltage value. Replicating the process on images recorded at discrete voltage values then provides bias-voltage-dependent current data (*I-V*) for all defects in a scan frame. Polynomial fits to the *I-V* data can then be produced, the slopes of which are subsequently calculated to extract defect-specific d*I*/d*V-V* information.

Following the procedure described above, we performed *I-V* and d*I*/d*V-V* analysis on all defects observed in Fig. 2(c, d). This was done by way of images collected between bias voltages of +2 V and -2 V, in 0.5 V increments (with 5 images per bias voltage value), the results of which are presented in Fig. 3. Studying the *I-V* and d*I*/d*V-V* behavior summarized in Fig. 3, we can categorize the observed defects into three different groups. Defects 1 and 2 (Group 1) weakly enhance the conductivity at both positive and negative bias voltages, with a relatively flat DOS profile across the voltage range. On the other hand, defects 3, 4, and 6 (Group 2) exhibit a prominent increase in conductivity with increasing bias voltage in the positive region, accompanied by relatively weak and bias-independent enhancement of current at bias voltages of -0.5 V and below. Defects 5 and 7 form the last set (Group 3), with their characteristic behavior being a prominent increase in DOS at increasingly negative bias voltages, and otherwise similar behavior to Group 1 defects.

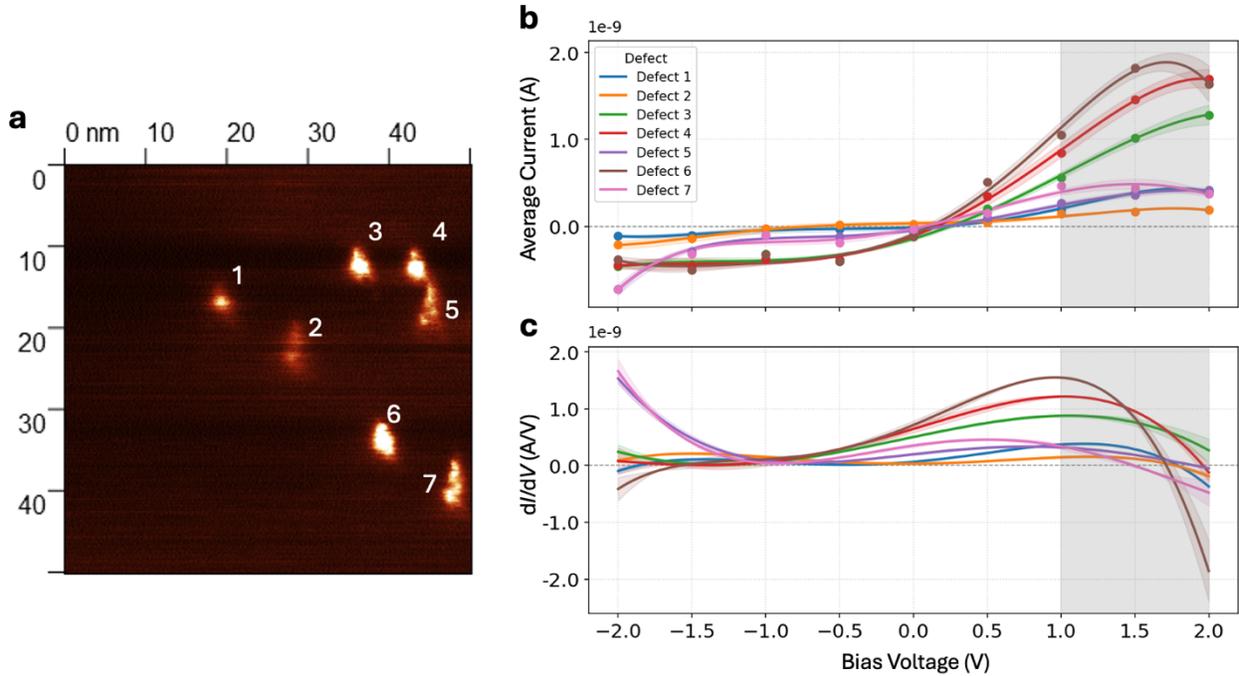

**FIG 3.** (a) Current map of defects on MoS$_2$ imaged at a bias voltage of +2 V. (b) Average current values are plotted as a function of the bias voltage, for each of the seven defects observed in (a). Each set of points is fitted with a 5$^{th}$ degree polynomial, resulting in a characteristic, continuous *I-V* curve. (c) The derivative of each fitted curve from (b), producing the d*I*/d*V-V* curve for all observed defects in (a). The shaded area in (b) and (c) highlights the potential occurrence of a tip change, causing an unexpected drop in slope.

It should be noted that the relatively sparse sampling of the voltage range in the experiments results in individual data points having significant influence on fitted curves. In particular, an unexpected drop in current slopes happens for the majority of defects at/around a bias voltage of +1 V (shaded region in Fig. 3(c), in some cases into the negative regime), which potentially points to a gradual change in the tip-sample contact. On the other hand, other features such as a prominent increase in current at increasingly negative bias voltages only happens for a certain set of defects (i.e., Group 3), allowing their robust assignment as defect-specific changes in electronic properties.

It is particularly important to highlight that the grouping of defects performed in this fashion is strongly supported by the morphological features observable from current maps. Specifically, the spatial shape of defects assigned to each group are consistent with each other, as one can see in the current map presented in Fig. 3(a) (Defects 1 and 2; Defects 3, 4 and 6; Defects 5 and 7). This provides further validation for the applicability of the discrete *I-V* methodology for defect categorization.

Comparison of electronic spectroscopy results with existing studies in the literature allows making conclusions about the chemical identity of the defects. In particular, the relatively flat d$I$/d$V$ plots approaching 0 for the majority of defects in the negative voltage regime (except for Group 3) is indicative of defect-induced band bending or charge trapping mechanisms that were previously observed for transition metal substitutions in STM and C-AFM studies of MoS$_2$.[4] Another key piece of information regarding the chemical identity of the defects is gathered by comparing their character in terms of current-enhancing vs. current-attenuating at opposing bias voltage polarities. In particular, all defects observed in Fig. 3 enhance current locally for both positive and negative bias voltages. This polarity-independent, current-enhancing character exhibited by defects support their assignment as non-charged (i.e. neutral) transition metal substitutions, in alignment with studies performed by Bampoulis *et al.*[19] and Dunn *et al.*[18] While additional factors such as the lateral extent (several nanometers) and the spatial density (~1.5-3.0×10$^{11}$ cm$^{-2}$) of defects observed in Figs. 2 and 3 further support their general assignment as transition metal substitutions,[19] additional details in the d$I$/d$V$-$V$ data allow more distinctions to be made between the defects. Specifically, defects in Group 3 exhibit enhanced current and DOS at increasingly negative bias voltages, as opposed to others. This behavior points to a local abundance of donor states, and enables their identification as n-type transition metal dopants (such as Re).[20] On the other hand, the significant increase in current and DOS at increasingly positive bias voltages for Group 2 defects is indicative of strong p-type doping that would be induced by dopants such as V or Nb.[21] Finally, the effect of Group 1 defects on the DOS is less noticeable than Group 2 and 3, pointing towards a different transition metal dopant than Group 2 and 3 defects.

The discrete $I$-$V$ spectroscopy approach described here can also be extended to the study of point (i.e., single-atomic) defects in order to elucidate their bias-voltage-dependent behavior and consequently, their chemical identities. The results of such an effort are presented in Fig. 4, where two point defects are detected in a small-scale ($10 \times 10$ nm$^2$) current scan on MoS$_2$. The defects are locally attenuating the current in comparison to the surrounding atomic lattice, at both bias polarities. Recording the currents on a particular defect (highlighted by the dashed black circle in Fig. 4(a)) as a function of bias voltage in a range of -1.25 V to +1.25 V then allows the extraction of characteristic $I$-$V$ behavior (Fig. 4(b)). Comparison of $I$-$V$ behavior with a non-defective location in the vicinity of the defect reveals close overlap in the positive voltage range (i.e., towards the conduction band), whereby the current detected over the negative voltage range (i.e., towards the valence band) slightly deviates from the non-defective region. Moreover, no significant distortions (e.g., in the form of in-gap states) are caused by the point defect in differential conductance (i.e., d$I$/d$V$-$V$) measurements (Fig. 4(c)).

At this point, it should be noted that there is considerable debate in the literature regarding the identity of point defects observed in TMDs. While S vacancies have been generally considered as the primary candidates for point defects in samples grown via chemical vapor deposition,[22] recent STM/STS and DFT work have pointed toward O atoms substituting S sites at or near the surface as the predominant point defect in flux-grown TMDs.[5] Unlike O substitutions, S vacancies are expected to result in in-gap states close to the conduction band minimum in $MoS_2$.[23] On the other hand, O substitutions do not induce in-gap states and $dI/dV$-$V$ spectra remain mostly unperturbed with slight modifications near the valence band maximum.[5, 11] These characteristics are in strong alignment with the data presented in Fig. 4. Taken together with the calculated density of point defects (~$2.0 \times 10^{12}$ $cm^{-2}$, which is in accordance with literauture,[24] and an order of magnitude higher than the density of transition metal substitutions reported in Fig. 3) and their current-attenuating character, these results enable their identification in our experiments as O substitutions. On the other hand, the data presented here do not point to a complete absence of S vacancies in $MoS_2$, as point defects inducing in-gap states have been experimentally observed in this material previously.[8]

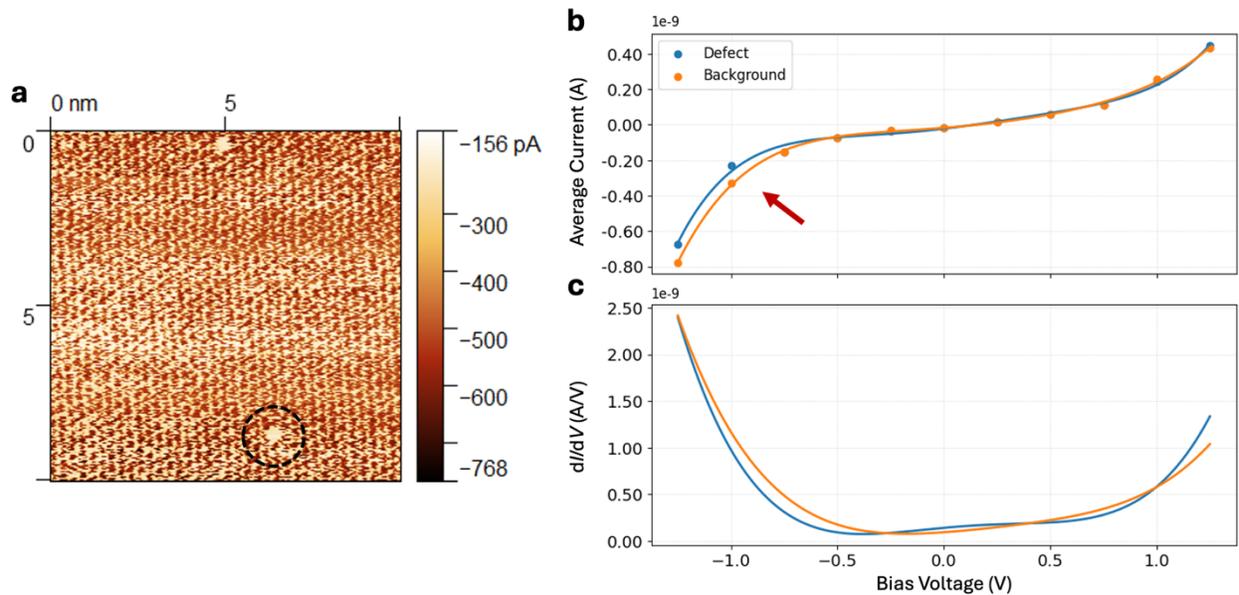

**FIG 4.** (a) A $10 \times 10$ $nm^2$ current map on $MoS_2$ imaged at a bias voltage of -1 V, where two point defects are detected. (b) Average current plotted as a function of the bias voltage for the point defect highlighted by the dashed black circle in (a) (blue), as well as a non-defective (i.e., background) region in its vicinity (orange). The two sets of points are fitted with 5th degree

polynomials. A deviation of current on the defect from the background signal is detected at increasingly negative voltages (red arrow). (c) The derivatives of the fitted curves from (b), in the form of d$I$/d$V$-$V$ curves for the defect (blue) and the background (orange).

We presented here a methodology termed "discrete $I$-$V$ spectroscopy" that enables the extraction of local electronic characteristics associated with individual atomic defects in the prototypical TMD, MoS$_2$. Our methodology notably works under ambient conditions with virtually no sample preparation, and thus much higher throughput than UHV-based methods such as STM. The data additionally allow the identification of defects based on signatures in local electronic behavior. The method described here is not restricted to TMDs and can be expanded to the study of other electronic and quantum materials under application-relevant (i.e., ambient) conditions.

**ACKNOWLEDGEMENTS**

This work was supported by the VISION (Venture for Innovation in Self-assembly and Integration of Optoelectronic Nanostructures) center at the University of California Merced, which is funded by the Partnerships for Research and Education in Materials (PREM) program of the National Science Foundation (NSF) via Award No. 2425230.

## REFERENCES

(1) Novoselov, K. S.; Geim, A. K.; Morozov, S. V.; Jiang, D.; Zhang, Y.; Dubonos, S. V.; Grigorieva, I. V.; Firsov, A. A. Electric field effect in atomically thin carbon films. *Science* **2004**, *306* (5696), 666-669. DOI: 10.1126/science.1102896.

(2) Manzeli, S.; Ovchinnikov, D.; Pasquier, D.; Yazyev, O.; Kis, A. 2D transition metal dichalcogenides. *Nature Reviews Materials* **2017**, *2* (8), 17033. DOI: 10.1038/natrevmats.2017.33.

(3) Krasheninnikov, A.; Batzill, M.; Delenda, A.; Drndic, M.; Ewels, C.; Franke, K.; Ghorbani-Asl, M.; Holleitner, A.; Jorio, A.; Kaiser, U.; et al. Defects and defect-mediated engineering of two-dimensional materials: challenges and open questions. *Beilstein Journal of Nanotechnology* **2026**, *17*, 454-488. DOI: 10.3762/bjnano.17.31.

(4) Addou, R.; Colombo, L.; Wallace, R. M. Surface Defects on Natural $MoS_2$. *ACS Applied Materials & Interfaces* **2015**, *7* (22), 11921-11929. DOI: 10.1021/acsami.5b01778.

(5) Holbrook, M.; Holtzman, L.; Hou, B.; Nashabeh, L.; Qiu, D.; Barmak, K.; Pasupathy, A.; Hone, J. Revealing Substitutional Oxygen as the Dominant Defect in Flux-Grown Transition Metal Diselenides. *Nano Letters* **2025**, *25* (37), 13795-13801. DOI: 10.1021/acs.nanolett.5c03126.

(6) Sun, J.; Passacantando, M.; Palummo, M.; Nardone, M.; Kaasbjerg, K.; Grillo, A.; Di Bartolomeo, A.; Caridad, J.; Camilli, L. Impact of Impurities on the Electrical Conduction of Anisotropic Two-Dimensional Materials. *Physical Review Applied* **2020**, *13* (4), 044063. DOI: 10.1103/PhysRevApplied.13.044063.

(7) Liang, Q.; Zhang, Q.; Zhao, X.; Liu, M.; Wee, A. Defect Engineering of Two-Dimensional Transition-Metal Dichalcogenides: Applications, Challenges, and Opportunities. *ACS Nano* **2021**, *15* (2), 2165-2181. DOI: 10.1021/acsnano.0c09666.

(8) Kumral, B.; Barri, N.; Demingos, P.; Adabasi, G.; Grishko, A.; Wang, G.; Kawase, J.; Onodera, M.; Machida, T.; Baykara, M.; et al. Mechanically reliable and electronically uniform monolayer $MoS_2$ by passivation and defect healing. *Nature Communications* **2025**, *16* (1), 7105. DOI: 10.1038/s41467-025-62370-0.

(9) Yang, Y.; Xu, K.; Pena, T.; Neilson, K.; Zheng, X.; Hoang, A.; Yang, K.; Hennighausen, Z.; Zhang, T.; Holtzman, L.; et al. Nondestructive Atomic Defect Quantification of Two-Dimensional Materials and Devices. *ACS Applied Materials & Interfaces* **2026**, *18* (6), 10161-10170. DOI: 10.1021/acsami.5c19328.

(10) Liu, S.; Liu, Y.; Holtzman, L.; Li, B.; Holbrook, M.; Pack, J.; Taniguchi, T.; Watanabe, K.; Dean, C.; Pasupathy, A.; et al. Two-Step Flux Synthesis of Ultrapure Transition-Metal Dichalcogenides. *ACS Nano* **2023**, *17* (17), 16587-16596. DOI: 10.1021/acsnano.3c02511.

(11) Barja, S.; Refaely-Abramson, S.; Schuler, B.; Qiu, D.; Pulkin, A.; Wickenburg, S.; Ryu, H.; Ugeda, M.; Kastl, C.; Chen, C.; et al. Identifying substitutional oxygen as a prolific point defect in monolayer transition metal dichalcogenides. *Nature Communications* **2019**, *10*, 3382. DOI: 10.1038/s41467-019-11342-2.

(12) Sumaiya, S. A.; Liu, J.; Baykara, M. Z. True Atomic-Resolution Surface Imaging and Manipulation under Ambient Conditions via Conductive Atomic Force Microscopy. *ACS Nano* **2022**, *16* (12), 20086-20093. DOI: 10.1021/acsnano.2c08321.
15

(13) Sumaiya, S.; Baykara, M. Atomic-scale imaging and spectroscopy via scanning probe microscopy: An overview. *Journal of Vacuum Science & Technology B* **2023**, *41* (6), 060802. DOI: 10.1116/6.0002889.

(14) Xu, K.; Holbrook, M.; Holtzman, L.; Pasupathy, A.; Barmak, K.; Hone, J.; Rosenberger, M. Validating the Use of Conductive Atomic Force Microscopy for Defect Quantification in 2D Materials. *ACS Nano* **2023**, *17* (24), 24743-24752. DOI: 10.1021/acsnano.3c05056.

(15) Adabasi, G.; Kumar, S.; Okay, E.; Evans, J.; Atli, E.; Ancheta, J.; Buke, G.; Martini, A.; Baykara, M. Strain-Modulated Conductivity and Work Function on Thin Crystals of $Mo_2C$. *ACS Applied Nano Materials* **2025**, *8* (41), 19810-19817. DOI: 10.1021/acsanm.5c03237.

(16) Adabasi, G.; Baykara, M. Scanning probe microscopy of MXenes: state of the art and prospects. *Nano Futures* **2025**, *9* (3), 032501. DOI: 10.1088/2399-1984/adea25.

(17) Yuan, Y.; Lanza, M. The Effect of Relative Humidity in Conductive Atomic Force Microscopy. *Advanced Materials* **2024**, *36* (51), 2405932. DOI: 10.1002/adma.202405932.

(18) Dunn, E.; Robson, A.; Young, R.; Jarvis, S. Single atom chemical identification of TMD defects in ambient conditions. *Nanotechnology* **2026**, *37* (13), 135705. DOI: 10.1088/1361-6528/ae5194.

(19) Bampoulis, P.; van Bremen, R.; Yao, Q.; Poelsema, B.; Zandvliet, H.; Sotthewes, K. Defect Dominated Charge Transport and Fermi Level Pinning in $MoS_2$/Metal Contacts. *ACS Applied Materials & Interfaces* **2017**, *9* (22), 19278-19286. DOI: 10.1021/acsami.7b02739.

(20) Murata, H.; Kataoka, K.; Koma, A. Scanning tunneling microscope images of locally modulated structures in layered materials, $MoS_2$(0001) and $MoSe_2$(0001), induced by impurity atoms. *Surface Science* **2001**, *478* (3), 131-144. DOI: 10.1016/S0039-6028(01)00904-9.

(21) Lu, S.; Leburton, J. Electronic structures of defects and magnetic impurities in $MoS_2$ monolayers. *Nanoscale Research Letters* **2014**, *9*, 676. DOI: 10.1186/1556-276X-9-676.

(22) Hong, J.; Hu, Z.; Probert, M.; Li, K.; Lv, D.; Yang, X.; Gu, L.; Mao, N.; Feng, Q.; Xie, L.; et al. Exploring atomic defects in molybdenum disulphide monolayers. *Nature Communications* **2015**, *6*, 6293. DOI: 10.1038/ncomms7293.

(23) Santosh, K.; Longo, R.; Addou, R.; Wallace, R.; Cho, K. Impact of intrinsic atomic defects on the electronic structure of $MoS_2$ monolayers. *Nanotechnology* **2014**, *25* (37), 375703. DOI: 10.1088/0957-4484/25/37/375703.

(24) Trainer, D.; Nieminen, J.; Bobba, F.; Wang, B.; Xi, X.; Bansil, A.; Iavarone, M. Visualization of defect induced in-gap states in monolayer $MoS_2$. *NPJ 2D Materials and Applications* **2022**, *6* (1), 13. DOI: 10.1038/s41699-022-00286-9.